# Classical Maxwellian polarization entanglement

John E. Carroll

*Centre for Photonic Systems*, University of Cambridge,
*9, J.J. Thomson Avenue, Cambridge, CB3 0FA, UK.*

E-mail: jec1000@cam.ac.uk

**Abstract.**

An explanation of polarization entanglement is presented using Maxwell's classical electromagnetic theory. Two key features are required to understand these classical origins. The first is that all waves diffract and weakly diffracting waves, with a principal direction of propagation in the laboratory frame, travel along that direction at speeds ever so slightly less than *c*. This allows non-trivial Lorentz transformations that can act on selected forward (*F*) waves or selected waves (*R*) travelling in the opposite direction to show that both can arise from a single zero momentum frame where all the waves are transverse to the original principal direction. Such *F* and *R* waves then both belong to a single relativistic entity where correlations between the two are unremarkable. The second feature requires the avoidance of using the Coulomb gauge. Waves, tending to plane waves in the limit of zero diffraction, can then be shown to be composed of two coupled sets of E and B fields that demonstrate the classical entanglement of *F* and *R* waves. Being derived from Maxwell's equations, the theory is compatible with special relativity. This is used to account for entanglement between waves travelling at arbitrary angles from a source. In spite of a classical explanation, selection of appropriate *F* and *R* waves means entanglement is likely to remain a quantum phenomenon.

## 1. Introduction

In a conference paper Carroll and Quarterman [1] argued that Maxwell's classical equations gave an explanation of polarization entanglement [2, 3, 4, 5]. Such a claim has also been made by other authors [6] using very different arguments based around the Maxwellian fluid model. The reality of entanglement has been irrefutably proved in recent experiments demonstrating correlated electron spins [7] so that a reasonably straightforward *classical* explanation from Maxwell's vector field equations, albeit for correlated photon polarizations, is of particular current interest. A key feature used in reference 1 and also used here is that weakly diffracting waves travel along a principle direction at velocities ever so slightly less than *c*. However the complications of 'adjoint' waves, introduced in that earlier work, are avoided in the present paper. The paper then goes further than reference 1 and demonstrates how this classical theory gives the maximum quantum mechanical violation for the CHSH (Clauser, Horne, Shimony,Holt) [8, 9] form of the Bell inequality[10] . Because Maxwell's equations satisfy special relativity, prima facie this entanglement theory satisfies special relativity. The argument is also helped by using those elements of Space Time Algebra (STA) [11, 12] that facilitate, with a minimal amount of algebra, an understanding of the effects of rotating the instruments that measure polarization.

The first step is to realize that, in spite of the frequent use of plane waves when explaining the quantization of Maxwell's equations [13, 14], there is no such entity as a wave that is uniform across the whole of space. It is claimed that plane waves, also known as TEM (Transverse Electric and Magnetic) waves [15, 16], should be considered as the diffraction free limit of TE (Transverse Electric) and TM (Transverse Magnetic) waves. There is of course a further objection that TE, TM and TEM waves have infinite energy [17] but, never the less, waves with finite energy can be broken down, through Fourier analysis, to a superposition of plane waves considered here as limiting cases of TE and TM fields when the diffraction and the axial fields tend to zero.

The next step then is to set up a model of weakly diffracting waves with a principle direction of propagation. Such waves travel along their principal direction at velocities strictly *less than c* in free space, even if by an immeasurably small amount. Consequently, there are always *finite* Lorentz transformations giving non-trivial boosts along the principal direction of propagation. These can change the laboratory reference frame containing the diffracting forward (*F*) wave, or alternatively containing the diffracting reverse (*R*) wave, to a 'proper' or 'zero momentum' frame of reference where all the wave motion is transverse to the selected principal propagation direction. Both *F* and *R* waves are then just different representations of the wave in that frame of reference with zero axial momentum. This is the first counter argument to the concerns that entanglement violates relativity [18]. If *F* and *R* waves are just different representations of exactly the same zero momentum wave packet then all parts of that packet communicate with all other parts immediately and directly. Consequently there is then no contradiction between special relativity and correlations of *F* and *R* waves. The concept of zero momentum wave packets has been extensively discussed by Lekner [19, 20] but entanglement of forward and reverse components is not believed to have been discussed.

This work also shows that it is important to avoid using the popular Coulomb gauge [21] where for the spatial vector potential **A**, div **A** = 0. Although this gauge is commonly used [13, 14, 22, 23] when discussing the quantization of plane waves, its use has hidden those classical arguments that can demonstrate how entanglement arises. All four components of a 1-vector potential **A** are required in order to demonstrate from the classical Maxwell equations that the plane wave is the limit of *two independent* sets of E- and B- fields. One set of E- and B- fields (called the E-field set and given a subscript *e*) will be shown to be generated by an axial electric field and the second set of E- and B- fields (the H-field set and given a subscript *h*) are generated by an axial magnetic field. These are the TM fields and TE fields mentioned earlier with an axial E-field or an axial H-field along the principal direction of propagation [16, 17]. In the limit of zero diffraction, these axial fields attached to the TM and TE fields tend to zero but their two distinct and independent contributions to the TEM limit remain, no matter how weakly diffracting is the wave. The Maxwellian explanation of entanglement is now developed in detail.

## 2. Space Time Algebra formulation

### 2.1 Fundamentals with weak diffraction

The development of a normalized set of Maxwell's equations using Space Time Algebra (STA) is rehearsed here, following concepts explained by Doran and Lasenby [12] but with notation used in the present work. The value of this analysis comes when spatially rotating any measurement instruments about the principal axis of propagation and also when considering boosts along that principal axis. The space-time axes are represented by unit anti-commuting orthogonal 1-vectors $\mathbf{e}_\mu$ where $\mathbf{e}_\mu \mathbf{e}_\nu = -\mathbf{e}_\nu \mathbf{e}_\mu$ ($\nu, \mu = 0,1,2,3$). The signature of time and space is taken here as $(1,-1,-1,-1)$ so that $(\mathbf{e}_0)^2 = 1$ ; $(\mathbf{e}_a)^2 = -1$ ($a = 1,2,3$). Given a 1-vector:

$$\mathbf{A} = A^0 \mathbf{e}_0 + A^1 \mathbf{e}_1 + A^2 \mathbf{e}_2 + A^3 \mathbf{e}_3, \tag{2.1}$$

then its space-time |magnitude|$^2$ is determined from the Lorentz invariant scalar:

$$(A^0\mathbf{e}_0 + A^1\mathbf{e}_1 + A^2\mathbf{e}_2 + A^3\mathbf{e}_3)^2 = (A^0)^2 - (A^1)^2 - (A^2)^2 - (A^3)^2. \tag{2.2}$$

Italic indices indicate coordinates while upright superscripts indicate powers. The pseudo-scalar multivector $\mathbf{I} = \mathbf{e}_0\mathbf{e}_1\mathbf{e}_2\mathbf{e}_3$ remains with the same sign when both space and time are reversed. Although $\mathbf{I}^2 = -1$, it must not to be confused with the imaginary scalar $i$ where one also has $i^2 = -1$. In this work, $i$ is used to allow cosine and sine to combine in a complex exponential. Without losing generality, the principal direction of propagation for a weakly diffracting wave can be taken in the laboratory reference frame as the O1 direction with its longitudinal coordinate $x^1$ and transverse coordinates $x^2$, $x^3$. For example with forward and reverse periodic waves, travelling principally along the direction O1, the transverse amplitudes can have the general form $A(\kappa x^2, \kappa x^3)$:

$$A_{Forward\,/\,Reverse} = A(\kappa x^2, \kappa x^3)\exp[-i(k_0 x^0 - /+ k_1 x^1)]. \tag{2.3}$$

Here $\kappa$ is an inverse length that gives a measure of the strength of the transverse diffraction with $\kappa$ tending to zero as the diffraction tends to zero. The normalized frequency is $k_0 = \omega/c$ and the normalized time is $x^0 = ct$ giving waves as in equations (2.3). By itself, $i$ changes a wave's phase by 90° but otherwise has none of the geometric properties exhibited by the multi-vector $\mathbf{I}$. For such waves in free space:

$$k_0^2 - k_1^2 = \kappa^2; \quad (\partial/\partial x^2)^2 + (\partial/\partial x^3)^2 = -\kappa^2. \tag{2.4}$$

The contravariant 1-vector electromagnetic potential is taken to be $\mathbf{A}$, as in equation (2.1). The covariant 1-vector 'world' differential is given the label □:

$$\square = (\partial_0\mathbf{e}_0 - \partial_1\mathbf{e}_1 - \partial_2\mathbf{e}_2 - \partial_3\mathbf{e}_3). \tag{2.5}$$

The negative signs are a consequence of $\partial/\partial(\mathbf{e}_a x^{(a)}) = -\mathbf{e}_a[\partial/\partial x^{(a)}] = -\mathbf{e}_a\partial_{(a)}$ : ($a=1$-$3$) while for the time coordinate $(1/\mathbf{e}_0)\,\partial/\partial(ct) = \mathbf{e}_0\partial_0$.

Taking the 'world' gradient of $\mathbf{A}$:

$$\square\,\mathbf{A} = (\partial_0\mathbf{e}_0 - \partial_1\mathbf{e}_1 - \partial_2\mathbf{e}_2 - \partial_3\mathbf{e}_3)(A^0\mathbf{e}_0 + A^1\mathbf{e}_1 + A^2\mathbf{e}_2 + A^3\mathbf{e}_3) \qquad (a)\;(2.6)$$

$$= (\partial_0 A^0 + \partial_1 A^1 + \partial_2 A^2 + \partial_3 A^3) \qquad (b)\;(2.6)$$

$$+ \sum_{j=1\text{-}3}(-\partial_0 A^j - \partial_j A^0)\,\mathbf{e}_j\mathbf{e}_0 - \sum_{k,\ell\,=2,3;\,3,1;\,1,2}\mathbf{e}_k\mathbf{e}_\ell(\partial_k A^\ell - \partial_\ell A^k). \qquad (c)\;(2.6)$$

The Lorentz invariant Lorenz condition [24, 25] sets the term in (2.6 b) to zero:

$$(\partial_0 A^0 + \partial_1 A^1 + \partial_2 A^2 + \partial_3 A^3) = 0. \tag{2.7}$$

The electric and magnetic bi-vectors $\mathbf{E}$ and $\mathbf{B}$ are given in terms (2.6 c) by:

$$\mathbf{E} = \sum_{j=1\text{-}3}\mathbf{e}_j\mathbf{e}_0 E^j = \sum_{j=1\text{-}3}(-\partial_0 A^j - \partial_j A^0)\,\mathbf{e}_j\mathbf{e}_0\,; \tag{2.8}$$

$$\mathbf{IB} = \sum_{j=1\text{-}3}\mathbf{e}_j\mathbf{e}_0\,\mathbf{I}B^j = -\sum_{k,\ell\,=2,3;\,3,1;\,1,2}\mathbf{e}_k\mathbf{e}_\ell(\partial_k A^\ell - \partial_\ell A^k). \tag{2.9}$$

The combination $\mathbf{\Psi} = \mathbf{E} + \mathbf{I\,B}$ is called the Riemann Silberstein bi-vector and is considered to be the quantum wave-function of the photon [26, 27]. In free space, the wave equation is given from:

$$\square^2\,\mathbf{A} = (\partial_0^2 - \partial_1^2 - \partial_2^2 - \partial_3^2)\,\mathbf{A} = \square\,\mathbf{\Psi}. \tag{2.10}$$

The wave equation is now partitioned to give (in free space):

$$(\partial_0^2 - \partial_1^2)\,A^\mu = -\kappa^2 A^\mu;\;\kappa^2 > 0\,;\,(\partial_2^2 + \partial_3^2)\,A^\mu = -\kappa^2 A^\mu. \tag{2.11}$$

Here, as already indicated, $\kappa$ is a simple single measure of the diffraction. The Lorenz condition is also partitioned to give:

$$(\partial_0 A^0 + \partial_1 A^1) = 0; \quad (\partial_2 A^2 + \partial_3 A^3) = 0. \tag{2.12}$$

These equations are satisfied by finding entities $\Theta$ and $\Phi$ (that will turn out to be the amplitudes of bi-vectors) along with an arbitrary scalar potential $\Upsilon$ such that:

$$A^0 = -\partial_1 \Theta + \partial_0 \Upsilon; \; A^1 = \partial_0 \Theta - \partial_1 \Upsilon; \; A^2 = \partial_3 \Phi - \partial_2 \Upsilon; \; A^3 = -\partial_2 \Phi - \partial_3 \Upsilon. \tag{2.13}$$

In free space:

$$\Box^2 \Theta = \Box^2 \Phi = \Box^2 \Upsilon = 0. \tag{2.14}$$

One may choose $\Upsilon$ to set $A^0 = 0$ giving, from equation (2.7), the 'Coulomb gauge' with div $\mathbf{A} = 0$. However this removes a set of transverse fields that are essential to a classical understanding of the phenomena of polarization entanglement. Here $\Upsilon$ is set to zero because its presence adds unnecessary algebra. The axial fields are then given from:

$$E^1 = -\partial_0 A^1 - \partial_1 A^0 = \kappa(\kappa\,\Theta); \quad B^1 = \partial_2 A^3 - \partial_3 A^2 = \kappa(\kappa\,\Phi). \tag{2.15}$$

The transverse fields are given from:

$$E^2 = -\partial_0 A^2 - \partial_2 A^0 = [\partial_1 (\partial_2/\kappa)(\kappa\,\Theta) - \partial_0 (\partial_3/\kappa)(\kappa\,\Phi)]; \tag{a) (2.16}$$

$$E^3 = -\partial_0 A^3 - \partial_3 A^0 = [\partial_1 (\partial_3/\kappa)(\kappa\,\Theta) + \partial_0 (\partial_2/\kappa)(\kappa\,\Phi)]; \tag{b) (2.16}$$

$$B^2 = \partial_3 A^1 - \partial_1 A^3 = [\partial_0 (\partial_3/\kappa)(\kappa\,\Theta) + \partial_1 (\partial_2/\kappa)(\kappa\,\Phi)]; \tag{c) (2.16}$$

$$B^3 = \partial_1 A^2 - \partial_2 A^1 = [-\partial_0 (\partial_2/\kappa)(\kappa\,\Theta) + \partial_1 (\partial_3/\kappa)(\kappa\,\Phi)]. \tag{d) (2.16}$$

Now as $\kappa \to 0$ it is assumed that in the diffraction free limit that:

$$(\kappa\,\Theta) \to \Theta_e; \; (\kappa\,\Phi) \to \Phi_h; \; (\partial_2/\kappa) \to D_2; \; (\partial_3/\kappa) \to D_3; \; D_2^2 + D_3^2 = -1. \tag{2.17}$$

Here although $\kappa \to 0$, $\Theta_e$ and $\Theta_h$ are always taken to have a non-zero finite limit. Similarly the normalized derivatives $D_{2/3}$ give non-zero finite outcomes. There are two different sets of fields (with $\kappa^2 > 0$ but arbitrarily small) created by $\Theta_e$ and $\Theta_h$. These sets have axial E and H fields respectively denoted by the subscripts $e/h$ as used above. Keeping for the present this arbitrarily small amount of diffraction:

$$(\partial_0^2 - \partial_1^2) = -\kappa^2; \quad D_2^2 + D_3^2 = -1. \tag{a) (2.18}$$

$$\mathbf{e}_1 \mathbf{e}_0 E^1 = \kappa\,\mathbf{e}_1 \mathbf{e}_0\,\Theta_e; \quad \mathbf{e}_1 \mathbf{e}_0\,\mathbf{I}\,B^1 = -\kappa\,\mathbf{e}_2 \mathbf{e}_3\,\Theta_h. \tag{b) (2.18}$$

This shows that $\Theta_e$ and $\Theta_h$ are proportional to the axial E- and B-fields and consequently, like the fields, are represented as bi-vectors [11, 12]. As $\kappa$ tends to zero so the axial E-and B-fields tend to zero while, as already mentioned, $\Theta_e$ and $\Theta_h$ remain finite. The transverse fields, still with a small but non-zero value of $\kappa$, are given from:

$$E^2 = \partial_1 D_2 \Theta_e - \partial_0 D_3 \Theta_h; \quad B^3 = -\partial_0 D_2 \Theta_e + \partial_1 D_3 \Theta_h; \tag{a) (2.19}$$

$$E^3 = \partial_1 D_3 \Theta_e + \partial_0 D_2 \Theta_h; \quad B^2 = \partial_0 D_3 \Theta_e + \partial_1 D_2 \Theta_h. \tag{b) (2.19}$$

Notice that as $\kappa \to 0$, $\partial_1 \to -\partial_0$, $E^1 \to 0$; $B^1 \to 0$; $E^2 \to B^3$ and $E^3 \to -B^2$: these form a classic right handed set of vector components for a plane wave propagating in the forward direction along O1. It may at first appear that there is nothing new.

To take advantage of STA in dealing with rotations and boosts, equations (2.19)(a) and (b) are re-written, with a bit of extra algebra, as bi-vector fields [12]:

$$\mathbf{e}_2\mathbf{e}_0 E^2 = (\mathbf{e}_1\partial_1)(\mathbf{e}_2 D_2)(\mathbf{e}_1\mathbf{e}_0\,\Theta_e) + (\mathbf{e}_0\partial_0)(\mathbf{e}_3 D_3)(\mathbf{e}_2\mathbf{e}_3\Theta_h); \qquad (a)\ (2.20)$$

$$\mathbf{e}_3\mathbf{e}_0\,IB^3 = -(\mathbf{e}_0\partial_0)(\mathbf{e}_2 D_2)(\mathbf{e}_1\mathbf{e}_0\,\Theta_e) + (\mathbf{e}_1\partial_1)(\mathbf{e}_3 D_3)(\mathbf{e}_2\mathbf{e}_3\,\Theta_h); \qquad (b)\ (2.20)$$

$$\mathbf{e}_3\mathbf{e}_0 E^3 = (\mathbf{e}_1\partial_1)(\mathbf{e}_3 D_3)(\mathbf{e}_1\mathbf{e}_0\,\Theta_e) + (\mathbf{e}_0\partial_0)(\mathbf{e}_2 D_2)(\mathbf{e}_2\mathbf{e}_3\,\Theta_h); \qquad (c)\ (2.20)$$

$$\mathbf{e}_2\mathbf{e}_0\,IB^2 = -(\mathbf{e}_0\partial_0)(\mathbf{e}_3 D_3)(\mathbf{e}_1\mathbf{e}_0\,\Theta_e) + (\mathbf{e}_1\partial_1)(\mathbf{e}_2 D_2)(\mathbf{e}_2\mathbf{e}_3\,\Theta_h). \qquad (d)\ (2.20)$$

Combine the bi-vector components and form the net 'E-' and 'B-' bivectors:

$$\mathbf{E} = (\mathbf{e}_1\partial_1)(\mathbf{e}_2 D_2 + \mathbf{e}_3 D_3)(\mathbf{e}_1\mathbf{e}_0\,\Theta_e)$$
$$+ (\mathbf{e}_0\partial_0)(\mathbf{e}_2 D_2 + \mathbf{e}_3 D_3)(\mathbf{e}_2\mathbf{e}_3\,\Theta_h); \qquad (a)\ (2.21)$$

$$\mathbf{IB} = -(\mathbf{e}_0\partial_0)(\mathbf{e}_2 D_2 + \mathbf{e}_3 D_3)(\mathbf{e}_1\mathbf{e}_0\,\Theta_e)$$
$$+ (\mathbf{e}_1\partial_1)(\mathbf{e}_2 D_2 + \mathbf{e}_3 D_3)(\mathbf{e}_2\mathbf{e}_3\,\Theta_h). \qquad (b)\ (2.21)$$

As $\kappa^2 \to 0$, equation (2.21) still holds with $\Theta_e$ and $\Theta_h$ tending to their finite limiting values, each giving *independent* sources of fields. As already noted, if $A^0 = 0$ (the Coulomb gauge) then $\Theta_e$ is removed and the analysis fails.

It is of interest to combine the bi-vectors in equations (2.21) to obtain the weak diffraction value of the Riemann Silberstein bi-vector:

$$\mathbf{\Psi} = \mathbf{E} + \mathbf{IB} = -\{(\mathbf{e}_0\partial_0 - \mathbf{e}_1\partial_1)_\kappa (\mathbf{e}_2 D_2 + \mathbf{e}_3 D_3)\,\mathbf{e}_1\mathbf{e}_0\Theta_e$$
$$- (\mathbf{e}_0\partial_0 + \mathbf{e}_1\partial_1)_\kappa (\mathbf{e}_2 D_2 + \mathbf{e}_3 D_3)\,\mathbf{e}_2\mathbf{e}_3\Theta_h\}. \qquad (2.22)$$

The subscript $\kappa$ has been attached to remind the reader that equation (2.22) holds with $\kappa^2 > 0$. As will be seen in section 2.4, this allows one to find a frame of reference where the boosted axial differential $\partial_1' \Rightarrow 0$.

## 2.2 Rotation about the principal axis

Rotations through an angle $\theta$ about the O1 axis are determined in STA by half angle rotors [11, 12] for example $R_{\frac{1}{2}\theta} = \exp(\frac{1}{2}\mathbf{e}_2\mathbf{e}_3\,\theta)$. Using the commutation properties of the rotor one can show that the bi-vector $\mathbf{\Psi}$ is then rotated into $\mathbf{\Psi}_\theta$:

$$\mathbf{\Psi}_\theta = (R_{\frac{1}{2}\theta})\,\mathbf{\Psi}\,(R_{\frac{1}{2}\theta})^{-1} \qquad (2.23)$$

If one keeps the same coordinates and rotates the relevant reference 1-vectors, then the only terms that change in equation (2.22) are

$$(\mathbf{e}_2 D_2 + \mathbf{e}_3 D_3) \Rightarrow (R_{\frac{1}{2}\theta})\,(\mathbf{e}_2 D_2 + \mathbf{e}_3 D_3)(R_{\frac{1}{2}\theta})^{-1}$$
$$= (\mathbf{e}_2' D_2 + \mathbf{e}_3' D_3) \qquad (2.24)$$

where

$$\mathbf{e}_2' = \mathbf{e}_2 \cos\theta + \mathbf{e}_3 \sin\theta;\ \mathbf{e}_3' = \mathbf{e}_3 \cos\theta - \mathbf{e}_2 \sin\theta;\ (\mathbf{e}_2'\mathbf{e}_3') = (\mathbf{e}_2\mathbf{e}_3). \qquad (2.25)$$

Alternatively the coordinates can change with the same reference 1-vectors so that the normalized differentials change to primed values:

$$(\mathbf{e}_2 D_2 + \mathbf{e}_3 D_3) \Rightarrow (\mathbf{e}_2 D_2' + \mathbf{e}_3 D_3'), \qquad (2.26)$$

where

$$D_2' = D_2 \cos\theta - D_3 \sin\theta;\ D_3' = D_3 \cos\theta + D_2 \sin\theta; \qquad (2.27)$$

When measuring $\mathbf{E}$ separately from $\mathbf{\Psi}$, as when measuring polarization, the rotational changes for $\mathbf{E}$ follow the same procedure. Equations (2.25) or (2.27) still hold, dependant on the required representation.

## 2.3 Boosts along the principal axis

Boosts with a rapidity $\beta$ along the O1 axis are also determined in STA by rotors [11,12] but now using a half value rapidity: $L_{\frac{1}{2}\beta} = \exp(-\frac{1}{2}\mathbf{e}_1\mathbf{e}_0\beta)$ :

$$\mathbf{\Psi}_\beta = (L_{\frac{1}{2}\beta})\mathbf{\Psi}(L_{\frac{1}{2}\beta})^{-1}. \tag{2.28}$$

If one keeps the same 1-vectors but boosts the relevant differentials, then the only terms that change in equation (2.22) are

$$(\mathbf{e}_0\partial_0 \pm \mathbf{e}_1\partial_1) \Rightarrow (L_{\frac{1}{2}\beta})(\mathbf{e}_0\partial_0 \pm \mathbf{e}_1\partial_1)(L_{\frac{1}{2}\beta})^{-1}$$
$$= (\mathbf{e}_0\partial_0' \pm \mathbf{e}_1\partial_1'),$$

where

$$\partial_0' = \partial_0 \cosh\beta + \partial_1 \sinh\beta; \quad \partial_1' = \partial_1 \cosh\beta + \partial_0 \sinh\beta. \tag{2.29}$$

The normalized transverse differentials $D_2$ and $D_3$ are unaltered by such a boost. The bi-vectors $(\mathbf{e}_1\mathbf{e}_0\Theta_e)$ and $(\mathbf{e}_3\mathbf{e}_2\Theta_h)$ are are also invariant for rotations in the plane $\mathbf{e}_2\mathbf{e}_3$ and similarly invariant to boosts along the direction $\mathbf{e}_1$.

## 2.4 'Proper' or 'zero momentum' frame of reference

For waves, as in equations (2.3), and satisfying equations (2.18)(*a*), one can find a value $\alpha$ for the forward (*F*) wave such that:

$$\partial_0 \Rightarrow -ik_0 = -i\,\kappa\cosh\alpha; \quad \partial_1 \Rightarrow ik_1 = i\,\kappa\sinh\alpha. \tag{2.30}$$

For the reverse (*R*) wave with the same magnitudes of $\kappa$ and $\alpha$:

$$\partial_0 \Rightarrow -ik_0 = -i\,\kappa\cosh\alpha; \quad \partial_1 \Rightarrow -ik_1 = -i\,\kappa\sinh\alpha. \tag{2.31}$$

Both *F* and *R* waves, given the correct launch conditions, can have identical magnitudes of $\kappa$ and $\alpha$. Using equations (2.29 - 2.31), the zero momentum frame where $\partial_1' = 0$ (the 'proper' frame for the weakly diffracting wave) is reached for the *F* wave by a boost $\beta = \alpha \rightarrow \partial_1' = 0$. Alternatively, for the *R* wave, *the same 'proper' frame* would be reached by a boost $\beta = -\alpha \rightarrow \partial_1' = 0$. Both *F* and *R* waves, if launched correctly, can therefore be formed from a single zero momentum wave through appropriate boosts along the O1 direction.

Of course as in equation (2.3), the transverse field patterns $A(\kappa x^2, \kappa x^3)$ of the *F* and *R* waves must satisfy the same transverse wave-equation and have the same diffraction parameter $\kappa$. In the limit as $\kappa \rightarrow 0$ the same transverse pattern still holds:

$$\lim_{\kappa \rightarrow 0}\{(1/\kappa^2)(\partial_2^2 + \partial_3^2)A(\kappa x^2, \kappa x^3)$$
$$= (D_2^2 + D_3^2)A(\kappa x^2, \kappa x^3) = -A(\kappa x^2, \kappa x^3)\}. \tag{2.32}$$

It is concluded that given precisely the right launch conditions from a single source, it is possible for the weakly diffracting *F* and *R* waves to have the same zero momentum fields where all the wave motion is entirely transverse to the original principal direction of propagation. In special relativity, entities *F* and *R* derived from a single frame through appropriate boosts, say $\beta_F$ and $\beta_R$, belong to the same relativistic entity. Then there is no mystery or violation of relativity if *F* and *R* are found to have correlated properties.

## 3. Polarization

### 3.1 Circular Polarization

Consider the polarization of the forward and reverse waves in more detail. Only the **E** bi-vector will be considered. From equation (2.21)*a*, **E** is re-written as:

$$\mathbf{E} = -(\mathbf{e}_0\partial_1)(\mathbf{e}_2 D_2 + \mathbf{e}_3 D_3)(\Theta_e) - (\mathbf{e}_0\partial_0)(\mathbf{e}_3 D_2 - \mathbf{e}_2 D_3)(\Theta_h).$$

$$\mathbf{E} = -(\mathbf{e}_0\partial_1)(\mathbf{e}_2 D_2 + \mathbf{e}_3 D_3)(\Theta_e) - (\mathbf{e}_0\partial_0)(\mathbf{e}_3 D_2 - \mathbf{e}_2 D_3)\Theta_h. \qquad (3.1)$$

As $\kappa \to 0$, $\partial_1 \to -\partial_0$ giving a limiting value of the **E** bi-vector for the *F* wave:

$$\mathbf{E}_F \Rightarrow (\mathbf{e}_0\partial_0)[\,\mathbf{e}_2(D_2\,\Theta_e + D_3\,\Theta_h) + \mathbf{e}_3(D_3\,\Theta_e - D_2\,\Theta_h)] \qquad (3.2)$$

As $\kappa \to 0$, $\partial_1 \to +\partial_0$ giving a limiting value of the **E** bi-vector for the *R* wave:

$$\mathbf{E}_R \Rightarrow (\mathbf{e}_0\partial_0)[\,\mathbf{e}_2(-D_2\,\Theta_e + D_3\,\Theta_h) - \mathbf{e}_3(D_3\,\Theta_e + D_2\,\Theta_h)] \qquad (3.3)$$

Circular polarization has transverse fields (say $\mathbf{E}_T$) where a rotation of 90° about the $\mathbf{e}_1$ axis (i.e. $\mathbf{e}_2\mathbf{e}_3\,\mathbf{E}_T$) followed by a ±90° phase change (i.e. $\pm i\,\mathbf{e}_2\mathbf{e}_3\,\mathbf{E}_T$) brings the field back to its original value $\mathbf{E}_T$: the + or – sign determines the sign of the polarization. One may then project out the two different circular polarizations using projection operators given by:

$$P_+ = \tfrac{1}{2}[1 + i\,\mathbf{e}_2\mathbf{e}_3];\; P_- = \tfrac{1}{2}[1 - i\,\mathbf{e}_2\mathbf{e}_3], \qquad (3.4)$$

with the properties:

$$P_+ P_+ = P_+\,;\; P_- P_- = P_-;\; P_+ P_- = P_- P_+ = 0.$$

$$P_{+/-}\,\mathbf{e}_2 = \tfrac{1}{2}[\,\mathbf{e}_2 +/-\, i\,\mathbf{e}_3];\; P_{+/-}\,\mathbf{e}_3 = \tfrac{1}{2}[\,\mathbf{e}_3 -/+\, i\,\mathbf{e}_2]; \qquad (3.5)$$

After some extra algebra, equations (3.2) and (3.3) can be rearranged:

$$P_+\,\mathbf{E}_F = \tfrac{1}{2}(\mathbf{e}_0\partial_0)(\mathbf{e}_2 + i\mathbf{e}_3)(D_2 - i\,D_3)(\Theta_e + i\,\Theta_h); \qquad \text{(a) (3.6)}$$

$$P_-\,\mathbf{E}_F = \tfrac{1}{2}(\mathbf{e}_0\partial_0)(\mathbf{e}_2 - i\mathbf{e}_3)(D_2 + i\,D_3)(\Theta_e - i\,\Theta_h); \qquad \text{(b) (3.6)}$$

$$P_+\,\mathbf{E}_R = -\tfrac{1}{2}(\mathbf{e}_0\partial_0)(\mathbf{e}_2 + i\mathbf{e}_3)(D_2 - i\,D_3)(\Theta_e - i\,\Theta_h); \qquad \text{(c) (3.6)}$$

$$P_-\,\mathbf{E}_R = -\tfrac{1}{2}(\mathbf{e}_0\partial_0)(\mathbf{e}_2 - i\mathbf{e}_3)(D_2 + i\,D_3)(\Theta_e + i\,\Theta_h); \qquad \text{(d) (3.6)}$$

The elegance of STA allows one to have an arbitrary rotation about the O1 axis for the detectors for either the *F* or *R* waves, by replacing

$$(\mathbf{e}_2 \pm i\mathbf{e}_3) \; with \; (\mathbf{e}_2' \pm i\mathbf{e}_3') \; or \; (\mathbf{e}_2'' \pm i\mathbf{e}_3''). \qquad (3.7)$$

Here the different primes indicate that the transverse axes have been rotated within their own plane by different rotors (rotating through different angles). The spatial direction of any E-field is unaffected by the term $\mathbf{e}_0\,\partial_0$ so that all the E-fields in this plane wave limit are orthogonal to the direction O1 as expected.

If the *F* wave is measured in a laboratory by, say, Felicia and found to have a 'positive' polarization with zero 'negative' polarization then from equation (3.6)(b) one argues that Felicia's measurement has selected:

$$(D_2 + i\,D_3)(\Theta_e - i\,\Theta_h) = 0 \Rightarrow$$
$$(D_2^{\,2} + D_3^{\,2})(\Theta_e - i\,\Theta_h) \Rightarrow -(\Theta_e - i\,\Theta_h) = 0 \qquad (3.8)$$

While $|\Theta_e| = |\Theta_h|$ might be expected from equipartition of energy in a single excitation process, $\Theta_e = i\Theta_h$ determines the type of polarization, independently of the rotational position of Felicia's equipment. Equation (3.8) along with (3.6)(*c*) and (3.7) show that if Felicia's partner, say Robert, attempts to measure circular polarization on the *R* wave then, again regardless of the transverse angle of his measuring equipment relative to Felicia's equipment, he must measure only negative polarization. Similarly if Felicia selects, in her measurement, negative polarization then $\Theta_e = -i\Theta_h$ and consequently Robert, provided he is also measuring circular polarization, can measure only positive polarization. Felicia and Robert always measure orthogonal circular polarizations provided that they measure these carefully selected waves from the identical single source.

*3.2 Linear Polarization*

Linear polarization may be considered in a similar way. The limiting E-bivectors for the *F* and *R* waves are found from equations (3.2) and (3.3) to give:

$\mathbf{E}_F \Rightarrow (\mathbf{e}_0 \partial_0)[\ \mathbf{e}_2(D_2\ \Theta_e + D_3\ \Theta_h) + \mathbf{e}_3(D_3\ \Theta_e - D_2\ \Theta_h)]$

$= \frac{1}{2}(\mathbf{e}_0 \partial_0)\{\ [(\mathbf{e}_2 + \mathbf{e}_3)\ D_2 - (\mathbf{e}_2 - \mathbf{e}_3)\ D_3](\Theta_e - \Theta_h)$
$\qquad + [(\mathbf{e}_2 + \mathbf{e}_3)\ D_3 + (\mathbf{e}_2 - \mathbf{e}_3)\ D_2](\Theta_e + \Theta_h);$ (3.9)

$\mathbf{E}_R \Rightarrow (\mathbf{e}_0 \partial_0)[\ \mathbf{e}_2(-D_2\ \Theta_e + D_3\ \Theta_h) - \mathbf{e}_3(D_3\ \Theta_e + D_2\ \Theta_h)]$

$= -\frac{1}{2}(\mathbf{e}_0 \partial_0)\{\ [(\mathbf{e}_2 + \mathbf{e}_3)\ D_2 - (\mathbf{e}_2 - \mathbf{e}_3)\ D_3](\Theta_e + \Theta_h)$
$\qquad + [(\mathbf{e}_2 + \mathbf{e}_3)\ D_3 + (\mathbf{e}_2 - \mathbf{e}_3)\ D_2](\Theta_e - \Theta_h);$ (3.10)

It is helpful now to note new orthogonal unit magnitude 1-vectors $\mathbf{e}_s$ and $\mathbf{e}_d$:

$(\sqrt{\frac{1}{2}})\ (\mathbf{e}_2 +/- \mathbf{e}_3) = \mathbf{e}_{s/d}\ ;\ \mathbf{e}_s^2 = \mathbf{e}_d^2 = -1;\quad \mathbf{e}_s\mathbf{e}_d + \mathbf{e}_d\mathbf{e}_s = 0.$ (3.11)

The 1-vectors $\mathbf{e}_s$ and $\mathbf{e}_d$ are then used to create yet further orthogonal unit magnitude 1-vectors $\mathbf{e}_a$ and $\mathbf{e}_b$:

$\mathbf{e}_a = \mathbf{e}_s\ D_2 - \mathbf{e}_d\ D_3;\ \mathbf{e}_b = \mathbf{e}_s\ D_3 + \mathbf{e}_d\ D_2\ ;$

$\qquad\qquad \mathbf{e}_a^2 = \mathbf{e}_b^2 = -1;\quad \mathbf{e}_a\mathbf{e}_b + \mathbf{e}_b\mathbf{e}_a = 0.$ (3.12)

Hence $\mathbf{e}_a$ and $\mathbf{e}_b$ may be used as the new basis 1-vectors for the transverse fields. These transverse 1-vectors rotate with rotors just like $\mathbf{e}_1$ and $\mathbf{e}_2$ so that one may write primes on the 1-vectors to denote *arbitrary* rotations of the transverse axes for the instrumentation of both Felicia and Robert. Initially we shall assume that both sets of measuring instruments are aligned together:

$\mathbf{E}_F \Rightarrow (\sqrt{\frac{1}{2}})(\mathbf{e}_0\partial_0)[\ \mathbf{e}_a'\ (\Theta_e - \Theta_h) + \mathbf{e}_b'\ (\Theta_e + \Theta_h)];$

$\qquad\mathbf{E}_R \Rightarrow -(\sqrt{\frac{1}{2}})(\mathbf{e}_0\partial_0)[\ \mathbf{e}_a'\ (\Theta_e + \Theta_h) + \mathbf{e}_b'\ (\Theta_e - \Theta_h)].$ (3.13)

## 4. The CHSH Theoretical Results from Maxwell

The equations (3.13) then reveal remarkable results. From the forward-reverse symmetry it cannot matter who makes the first measurement. Here it is assumed that it is Felicia. She makes a polarization measurement which finds, for example, that the direction of the E-field is aligned wholly along one particular transverse direction, say $\mathbf{e}_a'$. This direction will be arbitrarily labeled as a '+1' event for Felicia. The transverse direction $\mathbf{e}_b'$ is perpendicular to $\mathbf{e}_a'$ with no E-field measured along this $\mathbf{e}_b'$ direction, hence Felicia's measurement of linear polarization determines:

$\qquad(\Theta_e + \Theta_h) = 0.$ (4.1)

Until that first measurement of polarization, the relation between $\Theta_e$ and $\Theta_h$ is indeterminate.

The largest amplitude field with linear polarization that Robert can measure for the reverse travelling E-field, is given by the electric field that has to be consistent with equation (4.1). This direction is $\mathbf{e}_b'$ which is perpendicular to the direction measured by Felicia. This event where Felicia measures '+1' and Robert measures his optimum polarization will be called here a $N_{++}$ event.

Felicia, when using a two port linear polarizer, might just as well have found that she had measured a polarization aligned with $\mathbf{e}_b'$ at right angles to $\mathbf{e}_a'$; with $\Theta_e = \Theta_h$. This event is now said to be a '−1' event for Felicia. Robert's optimum measurement of linear polarization is now at right angles to $\mathbf{e}_b'$ whereas Felicia measures a polarization aligned with $\mathbf{e}_b'$. This different event where Felicia measures '−1' and Robert measures his new optimum polarisation will be called here a $N_{--}$ event.

When Felicia and Robert each have a two port linear polarizer then one may allow the orientation of Felicia's instrument to be at some angle $\varphi_a$ and the orientation of Roberts polarizer to be at some different angle $\varphi_b$ with respect to a common reference in a plane perpendicular to the principal axis. However it is only the relative difference angle $\varphi = \varphi_b - \varphi_a$ that is of relevance in this theory.

In general, if Robert is not able to communicate with Felicia, then Robert does not know what the optimum direction for his two port polarizer should be. Consequently he will find that $\varphi$ is an unknown transverse angle between his two port polarizer and Felicia's two port polarizer. When Felicia measures '+1', then Robert will have the E-fields with an amplitude $\cos \varphi$ along his '+1' ($N_{++}$ event) and an amplitude $\sin \varphi$ along his '−1' measurement direction ($N_{+-}$ event). When Felicia measures '−1', then Robert will have the E-fields with an amplitude $\cos \varphi$ along his '−1' measurement direction ($N_{--}$ event) and an amplitude $\sin \varphi$ along his '+1' measurement direction ($N_{-+}$ event).

Now use the usual classical/quantum rule that probabilities of measuring polarizations are proportional to |amplitude|$^2$. With an arbitrary angle $\varphi$, the probability of an $N_{++}$ event is $\cos^2 \varphi$ and similarly the probability of an $N_{--}$ event is $\cos^2 \varphi$. Both of these events are given a score of +1 (orthogonal polarizations measured by Felicia and Robert). The probabilities of the $N_{+-}$ and $N_{-+}$ events are both $\sin^2 \varphi$. Both of these events are given a score of −1 (parallel polarizations measured by Felicia and Robert). The expected probability of orthogonal polarizations for Felicia and Robert over many measurements (with an arbitrary $\varphi$) has then the 'CHSH probability' of:

$$\mathcal{E}(0, \varphi) = [(N_{++} + N_{--} - N_{+-} - N_{-+})/(N_{++} + N_{--} + N_{+-} + N_{-+})]$$
$$= [(\cos \varphi)^2 + (\cos \varphi)^2 - (\sin \varphi)^2 - (\sin \varphi)^2] / [(\cos \varphi)^2 + (\cos \varphi)^2 + (\sin \varphi)^2 + (\sin \varphi)^2]$$
$$= \cos 2\varphi. \qquad (4.2)$$

Accepting the Bell-CHSH inequalities [8, 10], we seek a range of values of $\varphi_b$ and $\varphi_a$ as follows:

$$S_{CHSH} = \mathcal{E}(\varphi_a, \varphi_b) - \mathcal{E}(\varphi_a, \varphi_b') + \mathcal{E}(\varphi_a', \varphi_b) + \mathcal{E}(\varphi_a', \varphi_b') \qquad (4.3)$$

'Preferred' angles [8] giving the maximum value of $S_{CHSH}$ are $\varphi_a = 0$, $\varphi_a' = \pi/4$, $\varphi_b = \pi/8$, $\varphi_b' = 3\pi/8$. However using the results of the theory in this section, it was noted that it is only the difference between the rotational angles of Felicia's and Robert's polarisers that matter. Consequently:

$S_{CHSH(max)} = \mathcal{E}(0, \pi/8) - \mathcal{E}(0, 3\pi/8) + \mathcal{E}(0, -\pi/8) + \mathcal{E}(0, \pi/8)$

$= \cos(\pi/4) - \cos(3\pi/4) + \cos(-\pi/4) + \cos(\pi/4) = (2\sqrt{2}) > 2.$ (4.4)

This is not a hidden variable theory [10] because the relative phases of $\Theta_e$ and $\Theta_h$ are not determined until the first measurement of polarization is made. It is only then that the relative polarization of the $F$ and $R$ waves are determined. The value $|S_{CHSH(max)}| = 2\sqrt{2}$ is the same value as predicted by quantum theory [8, 10] although it is Maxwell's classical equations that give this prediction with the standard classical/quantum assumption that |amplitude|$^2$ gives the probability of measuring polarization.

## 5. Conclusions

Two key features allow one to understand Maxwellian classical entanglement:

(1) Because weakly diffracting waves travel along their principal axis at speeds ever so slightly less than $c$, there is always a zero momentum frame or a 'proper' frame of reference that can be accessed using non-trivial Lorentz transformations. Forward ($F$) and reverse ($R$) waves can be created by appropriate non-trivial boosts of the waves contained in this single zero momentum frame of reference. In that case $F$ and $R$ waves are integral parts of a single relativistic entity and so are in immediate communication with one another without violating relativity.

(2) One must not use the Coulomb gauge for the potentials. All four space-time components of the vector potential are required to create the two independent sets of transverse fields essential for demonstrating the numerical properties of entanglement, required by the Bell inequalities, directly from Maxwell's classical equations. The algebra shows that the relative polarizations of $F$ and $R$ waves are determined only on the first polarization measurement of either the $F$ or $R$ wave.

There is of course an immediate problem with any classical explanation. Why is entanglement only observed in a quantum context? The answer here is that polarization entanglement is observed only if the weakly diffracting $F$ and $R$ waves have the same 'proper' or zero momentum frame of reference i.e. the two waves are generated by appropriate changes of rapidity from exactly the same zero momentum source. To have a fully classical result one would need to have every excitation of every photon to be in the same identical phase relationship with the same relationships between $\Theta_e$ and $\Theta_h$. This difficulty suggests that, for some time to come, entanglement is likely to remain a phenomenon of quantum measurements where it is easier, for example, to pair up photons from one source.

This is not a hidden variable solution [10] to the entanglement problem. The relative phases of the two independent field generators $\Theta_e$ and $\Theta_h$ are determined only when the first measurement is made by either Felicia or by Robert. It is this first measurement which determines the relative polarizations of the forward $F$ and reverse $R$ waves.

Practical experiments in general have some acute angle between the source S and the two observers F and R. In this idealized theory although FSR are all in a single line, the weakly diffracting waves travel in opposite directions along the directions SF and SR at a speeds $c-\delta$, where $|c-\delta|$ is less than $c$ if only by an immeasurably small amount. One can then always find finite Lorentz transformations at right angles to the line FR and so change the angle between S, F and R to an arbitrary acute angle. The fact that FSR is a straight line in this work is just a matter of convenience for the algebraic theory.

# References

[1] J. E. Carroll and A. H. Quarterman, "Relativistic Entanglement From Maxwell's Classical Equations". *The Physics Of Reality: Space, Time, Matter, Cosmos. Proceedings Of The 8th Symposium Honoring Mathematical Physicist Jean-Pierre Vigier*. Edited by R L. Amoroso, L. H. Kauffman and P. Rowlands. World Scientific Publishing Co. Pte. Ltd., ISBN #9789814504782, pp. 306-315 (2013).

[2] A. Aspect, P. Grangier, and G. Roger,"Experimental Realization of Einstein-Podolsky-Rosen-Bohm Gedankenexperiment: A New Violation of Bell's Inequalities", *Phys. Rev. Lett.* **49**, 91 (1982).

[3] Y. H. Shih and C. O. Alley, "New Type of Einstein-Podolsky-Rosen-Bohm Experiment Using Pairs of Light Quanta Produced by Optical Parametric Down Conversion", *Phys. Rev. Lett.* **61**, 2921 (1988).

[4] G. Weihs, T. Jennewein, C. Simon, H. Weinfurter, and A. Zeilinger, "Violation of Bell's Inequality under Strict Einstein Locality Conditions", *Phys. Rev. Lett.* **81**, 5039 (1998).

[5] M. A. Rowe, D. Kielpinski, V. Meyer, C. A. Sackett, W. M. Itano, C. Monroe and D. J. Wineland, "Experimental violation of a Bell's inequality with efficient detection", *Nature* **409**, 791-794 [doi:10.1038/35057215] (2001).

[6] R. Brady and R. Anderson, Maxwell's fluid model of magnetism, arXiv:1502.05926v1 [quant-ph] (2015).

[7] B. Hensen, H. Bernien, A.E. Dréau, A. Reiserer, N. Kalb, M.S. Blok, J. Ruitenberg, R.F.L. Vermeulen, R.N. Schouten, C. Abellán, W. Amaya, V. Pruneri, M. W. Mitchell, M. Markham, D.J. Twitchen, D. Elkouss, S. Wehner, T.H. Taminiau, and R. Hanson; "Experimental loophole-free violation of a Bell inequality using entangled electron spins separated by 1.3 km"  arXiv:1508.05949v1 (2015).

[8] J. F. Clauser, M. A. Horne, A. Shimony, and R. A. Holt, "Proposed Experiment to Test Local Hidden-Variable Theories", *Phys. Rev. Lett.* **23**, 880 –884 (1969); Erratum: Phys. Rev. Lett. 24, 549 (1970).

[9] N. Hoglund and O. Jacobson, "Bell's Theorem and Inequalities, with Experimental Considerations" *SA104X Degree Project in Engineering Physics, First Level* Department of Theoretical Physics, Royal Institute of Technology (KTH), http://www.diva-portal.org/smash/get/diva2:643754/FULLTEXT01.pdf  (2013).

[10] J. S. Bell "*Speakable and unspeakable in quantum mechanics*", CUP, Cambridge, ISBN 0521 36869 3   (1987).

[11] D. Hestenes, "*Space-Time Algebra*" Springer   ISBN 978-3-319-18413-5  (2015)

[12] C. Doran and A. Lasenby  "*Geometric Algebra for Physicist*", CUP, ISBN 0521 48022, (2003).

[13] D. A. B, Miller, "*Quantum Mechanics for Scientists and Engineers*", CUP, pp513, 521, ISBN: 9780521897839, (2008).

[14] D. Marcuse, "*Engineering Quantum Electrodynamics*",  pp, 58,68, Harcourt Brace & World Library of Congress No. 70-97860   (1970).

[15] J. D. Jackson, "*Classical Electrodynamics*", 2$^{nd}$ Ed. Wiley, New York pages 342 ISBN 10: 047143132X   (1975).

[16] S. Ramo, J. R.Whinnery and T. Van Duzer,  "*Fields and Waves in Communications Electronics*",  3$^{rd}$ Edition,  J,Wiley   New York     ISBN 10: 0471585513  (1994).

[17] J. Lekner,  "TM, TE and `TEM' beam modes: exact solutions and their problems", *J.Opt.A:Pure Appl.Opt.*, **3**,   407 doi:10.1088/1464-4258/3/5/314 (2001).

[18] D. Z Albert and R. Galchen, "Was Einstein Wrong?: A Quantum Threat to Special Relativity", *http://www.scientificamrican.com/article/was-einstein-wrong-about-relativity,* p186 (March 2009).

[19] J. Lekner,  "Electromagnetic pulses which have a zero momentum frame", *J. Opt. A: Pure Appl. Opt.* **5** L15 doi:10.1088/1464-4258/5/4/101 (2003).

[20] J. Lekner, "Angular momentum of electromagnetic pulses", *J. Opt. A: Pure Appl. Opt.* **6** S128 doi:10.1088/1464-4258/6/3/021 (2004)..

[21] J. D. Jackson, "*Classical Electrodynamics*", 2$^{nd}$ Ed. pages 220-223,Wiley, New York (1975).

[22] C. C. Gerry & P. L. Knight, "*Introductory Quantum Optics*" p18-20 CUP, ISBN 0 521 52735 X , (2005).

[23] R. Loudon, "*The Quantum Theory of Light*",  p127, 3rdEdition, Oxford Science Publications, ISBN0 19 850176 (2000).

[24] L. Lorenz "On the identity of the vibrations of light with electrical currents", *Philosophical Magazine,* Series 4, Vol**. 34**, Issue 230 · pages 287-301. DOI:10.1080/14786446708639882 . (1867).
[25] R. Nevels and C-S. Shin, "Lorenz, Lorentz, and the Gauge" *IEEE Antennas and Propagation Magazine,* **43**, No. 3, p70-71, http://www.engr.mun.ca/~egill/index_files/7811_w10/lorenz_gauge.pdf, (2001).
[26] I. Bialynicki-Birula and Z. Bialynicka-Birula, "The role of the Riemann-Silberstein vector in classical and quantum theories of electromagnetism", *J. Phys. A Math. Theor.* 11/2012; 46(15), DOI: 10.1088/1751-8113/46/5/053001 (2012).
[27] I. Bialynicki-Birula and Z. Bialynicka-Birula, "Corrigendum: The role of the Riemann–Silberstein vector in classical and quantum theories of electromagnetism", *J.Phys.A:Math.Theor.* 46053001, Online stacks.iop.org/JPhysA/46/15950 (2013).

## Acknowledgements

The author is grateful to Adrian Quarterman who helped to sow the seeds of thought [1] in the present paper and to Ian White for his continued support.